# Nigral diffusivity, but not free water, correlates with iron content in Parkinson's disease

**Jason Langley[1], Daniel E. Huddleston[2] and Xiaoping P. Hu[1,3]**

[1] Center for Advanced Neuroimaging, University of California Riverside, Riverside, CA, USA
[2] Department of Neurology, Emory University, Atlanta, GA, USA
[3] Department of Bioengineering, University of California Riverside, Riverside, CA, USA

E-mail: xhu ,.at.' engr.ucr.edu

**Abstract**

*Introduction:* The loss of melanized neurons in the substantia nigra pars compacta is a primary feature in Parkinson's disease. Iron deposition occurs in conjunction with this loss. Loss of nigral neurons should remove barriers for diffusion and increase diffusivity of water molecules in regions undergoing this loss. In metrics from single-compartment diffusion tensor imaging models, these changes should manifest as increases in mean diffusivity and reductions in fractional anisotropy as well as increases in the free water compartment in metrics derived from bi-compartment models. However, studies examining nigral diffusivity changes from Parkinson's disease with single-compartment models have yielded inconclusive results and emerging evidence in control subjects indicates that iron corrupts diffusivity metrics derived from single-compartment models. We aimed to examine Parkinson's disease-related changes in nigral iron and diffusion measures from single- and bi-compartment models as well as assess the effect of iron on these diffusion measures in two separate Parkinson's cohorts.

*Methods:* Iron-sensitive data and diffusion data were analyzed in two cohorts: First, a discovery cohort consisting of 71 participants (32 control participants and 39 Parkinson's disease participants) was examined. Second, an external validation cohort, obtained from the Parkinson's Progression Marker's Initiative, consisting of 110 participants (58 control participants and 52 Parkinson's disease participants) was examined. The effect of iron on diffusion measures from single- and bi-compartment models was assessed in both cohorts.

*Results:* Measures sensitive to the free water compartment (discovery cohort: $P=0.006$; external cohort: $P=0.01$) and iron content (discovery cohort: $P<0.001$; validation cohort: $P=0.02$) were found to increase in substantia nigra of the Parkinson's disease group in both cohorts. However, diffusion markers derived from the single-compartment model (i.e. mean diffusivity and fractional anisotropy) were not replicated across cohorts. Correlations were seen between single-compartment diffusion measures and iron markers in the discovery cohort (iron-mean diffusivity: $r=-0.400$, $P=0.006$) and validation cohort (iron-mean diffusivity: $r=-0.387$, $P=0.003$) but no correlation was observed between a measure from the bi-compartment model related to the free water compartment and iron markers in either cohort.

*Conclusion:* The variability of nigral diffusion metrics derived from the single-compartment model in Parkinson's disease may be attributed to competing influences of increased iron content, which tends to drive diffusivity down, and increases in the free water compartment, which tends to drive diffusivity up. In contrast to diffusion metrics derived from the single-compartment model, no relationship was seen between iron and the free water compartment in substantia nigra.

## 1. Introduction

Parkinson's disease is a progressive neurodegenerative disorder characterized by resting tremor, bradykinesia, rigidity, and gait impairment and affects approximately 1% of the population over the age of 60 years.[1] Depigmentation of melanized neurons in substantia nigra pars compacta (SNpc) is a primary feature of Parkinson's disease with SNpc undergoing extensive loss of melanized neurons in the early stages of the disease[2,3] and iron is deposited in SNpc concurrently with the loss of melaninzed neurons.[4-6] The loss of melanized neurons should affect SNpc microstructure by removing barriers to diffusion and these microstructural differences can be examined *in vivo* with diffusion imaging, which measures the movement of molecular water.[7]

Diffusion tensor imaging (DTI) is sensitive to the diffusivity of water and allows researchers to probe the directionality of diffusion, known as fractional anisotropy (FA), and the extent of diffusion, known as mean diffusivity (MD), in tissue. Several studies have attempted to use DTI to characterize the effects of Parkinson's disease on SNpc microstructure.[8-14] However, no consensus has emerged from results in prior studies where lower nigral FA[10-13, 15] or no difference in nigral FA[8, 16-19] have been reported. This inconsistency has hampered the search for diagnostic and progression imaging markers of Parkinson's disease.[17]

While the use of standard (single-compartment) DTI has not yielded reproducible diagnostic biomarkers in SNpc, multi-compartment modeling approaches have yielded consistent diagnostic imaging markers of Parkinson's disease. In particular, the application of a bi-compartment model has reliably found increases in the CSF volume fraction (free water compartment) in SNpc of Parkinson's disease as compared to controls.[20-23] Other modeling approaches have revealed Parkinson's disease associated increases in SNpc kurtosis,[24] as well as a reduction in neurite density in Parkinson's disease using the three-compartment neurite orientation dispersion and density imaging model.[25]

Iron is deposited in SNpc after onset of Parkinson's disease.[11, 12, 26-33] Iron deposition in SNpc may alter diffusivity in the single-compartment model, contributing to the inconsistency of Parkinson's disease DTI imaging markers. Iron has been found to negatively bias diffusivity and positively bias FA in subcortical grey matter structures of older healthy adults.[34, 35] While emerging evidence indicates that iron and metrics from a single-compartment diffusion model are correlated, the effect of iron on metrics derived from multi-compartment models is unknown.

In this work, we use an atlas constructed from magnetization transfer effects, which are sensitive to neuromelanin,[36-39] to localize SNpc and examine Parkinson's disease-related microstructural changes in SNpc using single and multi-compartment models in two cohorts. We further examine the influence of Parkinson's disease-related microstructural and compositional changes on diffusion metrics with the goal of elucidating the variability observed in nigral microstructural measures from earlier studies examining microstructural changes in Parkinson's disease.

## 2. Materials and Methods

### 2.1 Subjects

Two cohorts were used in this analysis, a discovery cohort and a replication cohort. Ninety-two participants (52 participants with Parkinson's disease and 40 control participants) enrolled in the discovery cohort. Data from 13 Parkinson's disease and 8 control participants were excluded due to motion artifacts or incomplete scans. The final sample size for the discovery cohort was 71 subjects (39 Parkinson's disease participants and 32 control participants). Participants with Parkinson's disease were recruited from the Emory University Movement Disorders Clinic and clinically diagnosed with Parkinson's disease according to the Movement Disorders Society consensus diagnostic criteria.[40] Parkinson's disease patients had early to moderate disease with a Unified Parkinson's Disease Rating Scale Part III (UPDRS-III) ON medications motor score of ≤25. Control subjects were recruited from a cohort of individuals without major neurological diagnoses followed by the Emory Alzheimer's Disease Research Center. Specific exclusion criteria included the following: 1) patients showing symptoms or signs of secondary or atypical parkinsonism,[41] 2) controls were excluded if they scored ≤26 on the Montreal Cognitive Assessment (MOCA) indicating cognitive impairment, 3) any history of vascular territorial stroke, epilepsy, multiple sclerosis, neurodegenerative disease (aside from Parkinson's disease), peripheral neuropathy with motor deficits, parenchymal brain tumor, hydrocephalus, or schizophrenia, 4) treatment with an antipsychotic drug (other than quetiapine at a dose less than 200mg daily), or 5) if there were any contraindications to MRI imaging. All subjects participating in the discovery cohort gave written informed consent in accordance with local institutional review board regulations.

Demographic information including gender, age and education, was collected for each subject in the discovery cohort. Participants in both the Parkinson's disease and control groups underwent UPDRS-III examination by a fellowship-trained movement disorders neurologist. Parkinson's disease patients were examined and underwent imaging in the ON medication state.

A second cohort was obtained from the Parkinson's Progression Markers Initiative (PPMI) database (www.ppmi-info.org/data). PPMI is a multi-site study collecting imaging, biomarker, and clinical assessments in a group of *de novo* Parkinson's disease patients. For up-to-date information on the study, visit www.ppmi-info.org. Full inclusion and exclusion criteria for enrollment in PPMI can be found at the PPMI website (www.ppmi-info.org). Institutional IRB approved the study for each site and subjects gave written informed consent. Criteria for inclusion for subjects from the PPMI database used in this analysis were as follows: 1) participants must be scanned with cardiac-gated DTI and dual echo turbo-spin echo (TSE) acquisitions and 2) Parkinson's disease participants must have DTI and dual echo TSE scans at the 48-month time point with scan parameters matching those in the PPMI imaging protocol.

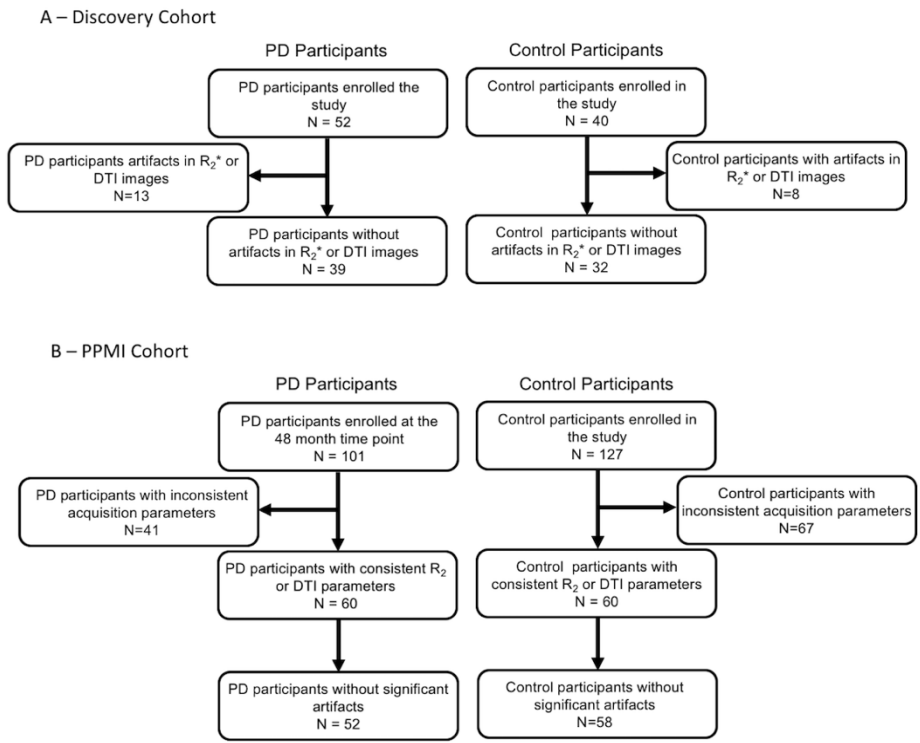

**Figure 1 Participant flow chart for both cohorts.** (**A**) Participant flow chart for the discovery cohort. (**B**) Participant flow chart for the PPMI cohort.

In the PPMI cohort, data from 228 participants (101 Parkinson's disease and 128 control participants) were downloaded in December of 2019. A total of 41 Parkinson's disease and 67 controls were excluded due to inconsistent scan parameters from the DTI or dual echo TSE acquisitions. After triage, 8 Parkinson's disease participants and 2 control participants were excluded due to motion artifacts in the DTI or dual echo TSE acquisitions. The final sample size for the PPMI cohort was 110 (58 control and 52 Parkinson's disease participants). A flow chart summarizing the calculation of the final sample size in both cohorts is shown in Figure 1.

*2.2 Image Acquisition*

Imaging data for the discovery cohort were acquired at Emory University on a 3T MRI scanner (Prisma Fit, Siemens Healthineers, Malvern, PA) using a 64 channel receive only coil. Images from a MP-RAGE sequence (echo time (TE)/repetition time (TR)/inversion time (TI)=3.02/2600/800 ms, flip angle=8°, voxel size=0.8×0.8×0.8 mm³) were used for registration from subject space to common space. Multiecho $T_2$*-weighted data were collected with a six echo 3D gradient recalled echo (GRE) sequence: $TE_1/\Delta TE/TR$=4.92/4.92/50 ms, FOV=220×220 mm², matrix size of 448×336×80, in-plane resolution = 0.49×0.49 mm², slice thickness=1 mm, and GRAPPA acceleration factor=2.

Diffusion MRI data in the discovery cohort were collected using a diffusion-weighted spin-echo EPI sequence with parameters matching the Lifespan Human Connectome Project: TE/TR=89.2/3222 ms, FOV=210×210 mm², matrix size=140×140, voxel size=1.5×1.5×1.5 mm³, partial Fourier=6/8, multiband factor=4, with 92 slices. Six images without diffusion weighting ($b$=0 sec/mm²) were also acquired with matching parameters. Monopolar diffusion encoding gradients were applied in 99 directions with $b$ values of 1000 s/mm² and 2000 s/mm². Two sets of diffusion-weighted images with phase-encoding directions of opposite polarity were acquired to correct for susceptibility distortion.[42]

$T_1$-weighted structural images in the PPMI cohort were acquired with a MP-RAGE sequence (TE/TR/TI=2.98/2300/900 ms, flip angle=9.0°, voxel size=1×1×1 mm³) were acquired for registration to common space. Dual echo turbo spin echo (TSE) images were acquired with the following parameters: $TE_1/TE_2/TR$=11/101/3270 ms; FOV=240×213 mm²; voxel size=0.9×0.9×3 mm³; 48 slices. Cardiac-gated diffusion-MRI data in the PPMI cohort were acquired using a monopolar diffusion encoding gradient with 64 unique gradient directions and the following parameters: TE/TR=88/650-1100 ms, flip angle=90.0°, FOV=229×229 mm², voxel size=1.98×1.98×2 mm³, $b$ value of 1000 s/mm², cardiac-triggered, with 72 slices.

*2.3 DTI Processing*

For both cohorts, diffusion data was preprocessed with FSL.[43-45] In the discovery cohort, magnetic field inhomogeneities were estimated from $b$=0 images with phase encoding gradients of opposite polarity. Susceptibility induced distortions, motion, and eddy-current induced distortions were corrected by EDDY. In the PPMI cohort, diffusion MR data were corrected for motion and eddy-current distortions using EDDY in FSL. Next, susceptibility distortions were reduced by nonlinearly fitting the $b=0$ image to second echo from the TSE acquisition. For both cohorts, parameters derived from the diffusion data were estimated using DTIFIT. A bi-compartment model,[46] implemented in DIPY,[47] was used to construct free water maps in both cohorts.

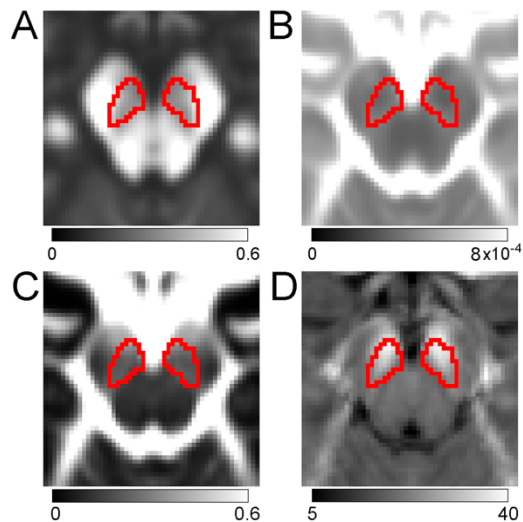

**Figure 2 Illustration of the SNpc ROI used in this analysis.** The SNpc ROI is outlined on mean SNpc FA (A), mean SNpc MD (B), and mean FW (C), and $R_2^*$ (D) images from the control group of the discovery cohort. For purposes of visualization, each participant's image (FA, MD, FW, and $R_2^*$) was transformed to MNI 1mm space and averaged.

For subjects in both cohorts, the $b=0$ image was brain extracted[48] and registered to the brain extracted $T_1$-weighted image using a rigid body transform with a boundary-based registration cost function.

### 2.3 Iron Processing

For both cohorts, $R_2^*$ (from multiecho GRE acquisition) or $R_2$ (from multiecho TSE acquisition) values were estimated using a custom script in MATLAB by fitting a monoexponential model to the multiecho GRE or multiecho TSE images. The resulting $R_2^*$ or $R_2$ maps were aligned to the $T_1$-weighted image using a rigid body transform derived via the magnitude image from the first echo.

### 2.3 Transformation of SNpc ROIs

A SNpc atlas, derived from magnetization-transfer images in a cohort of 76 healthy older participants (aged 66.6 years ± 6.4 years), was used as a region of interest (ROI) in the diffusion and iron analyses.[49] The SNpc atlas was transformed from MNI space to subject space using linear and nonlinear transforms in FSL[50, 51] as previously described.[34] Mean DTI and iron measures (discovery cohort: $R_2^*$, FA, MD, and free water (FW); PPMI cohort: $R_2$, FA, MD, and FW) were calculated in the SNpc for each subject. Figure 2 shows the SNpc ROI overlaid on mean SNpc FA, mean SNpc MD, and mean SNpc FW images from the control group in discovery cohort.

### 2.4 Statistical Analyses

Statistical analyses were performed using SPSS version 24 (IBM SPSS Statistics, Chicago, Illinois, USA). Quantitative data are expressed as mean ± standard deviation. Group differences in both cohorts were assessed using between-group t-tests. A one-tailed *t*-test with a significance level of $P=0.05$ was used for group comparisons of iron measures ($R_2^*$ and $R_2$) and we hypothesize an increase in mean SNpc iron measures will be observed in the Parkinson's disease group since histopathology and imaging studies in Parkinson's disease have noted an increase in SNpc iron.[6, 26, 52] A one-tailed *t*-test with a significance level of $P=0.05$ was used for group comparisons of diffusion measures from a single-compartment model (MD, RD, FA). We expect SNpc diffusivity will be reduced and SNpc FA will be increased in the Parkinson's disease group since earlier studies have noted that diffusivity (MD and RD) and FA are negatively and positively biased, respectively, by iron deposition.[34, 53] A one-tailed *t*-test with a significance level of $P=0.05$ was used for group comparisons of nigral free water measures since prior imaging studies found elevated free water in the substantia nigra of Parkinson's disease patients as compared to controls.[21, 22] We hypothesize that measures of free water will be increased in the Parkinson's disease group of both cohorts.

SNpc diffusivity and FA are correlated with iron.[34, 53] A *post hoc* analysis of covariance (ANCOVA) was performed to remove the influence of iron in these measures and evaluate Parkinson's disease-related microstructural changes in SNpc without contributions from iron. To assess the impact of Parkinson's disease severity on imaging biomarkers, clinical measures (UPDRS-III ON score and disease duration) were correlated with mean SNpc diffusion indices as well as with mean SNpc iron. A correlation was considered to be significant if $P<0.05$.

### 2.4 Data Availability

Data from the PPMI cohort can be obtained from the PPMI database (www.ppmi-info.org/data).

|  | Discovery Cohort | | | PPMI Cohort | | | |
|---|---|---|---|---|---|---|---|
| Variable | CO (n=32) | Parkinson's (n=39) | *P* Value | CO (n=58) | Parkinson's (n=52) | *P* Value | |
| Gender (M/F) | 12/20 | 20/19 | 0.07 | 40/18 | 33/19 | 0.21 | **Table 1.** Subject demographics and clinical information subjects used in both cohorts. |
| Age (yrs) | 65.5±9.2 | 63.7±10.2 | 0.41 | 61.7±10.9 | 63.2±9.9 | 0.45 | |
| Education (yrs) | 14.4±6.0 | 15.1±4.7 | 0.57 | 16.0±3.0 | 15.4±2.9 | 0.26 | |
| UPDRS-III ON score | 2.6±2.2 | 24.0±11.4 | $<10^{-4}$ | 0.8±1.3 | 18.5±12.6 | $<10^{-4}$ | |
| Disease Duration (yrs) | — | 4.0±3.6 | — | — | 4.6±0.6 | — | |

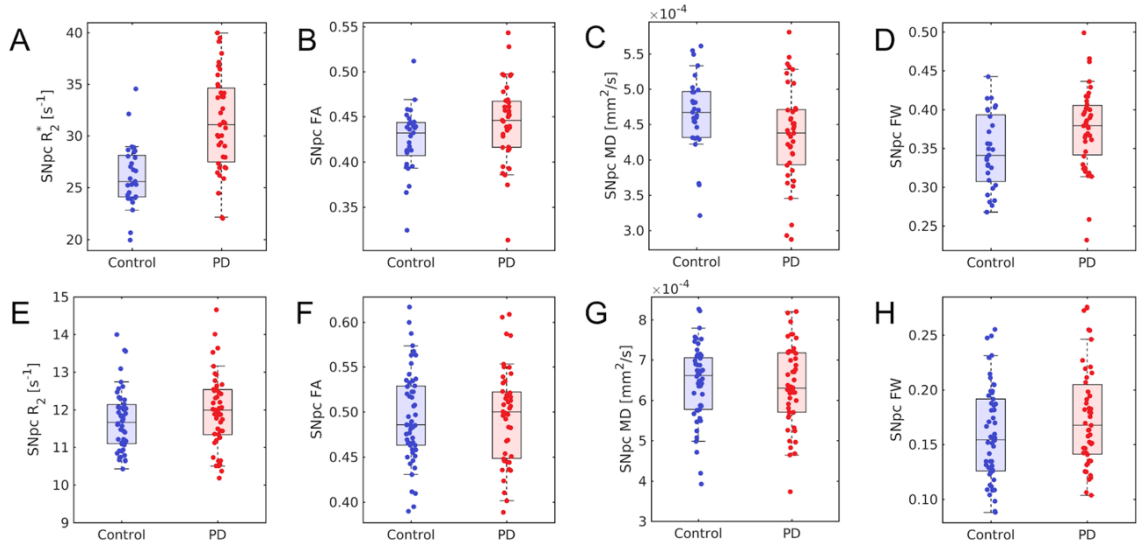

**Figure 3 Group comparisons for both cohorts**. Comparisons of $R_2^*$ (*P*<0.001), FA (*P*=0.001), MD (*P*=0.001), and FW (*P*=0.001) in the discovery cohort are shown in (**A-D**), respectively. The bottom row (**E-H**) show group comparisons for $R_2$ (*P*=0.023), FA (*P*=0.371), MD (*P*=0.330), and FW(*P*=0.01), respectively, in the PPMI cohort. In all box plots, the top and bottom of the box denote the 25th and 75th percentiles, respectively, with the line denoting the median value.

## 3. Results

No differences were observed in gender (discovery: *p*=0.07; PPMI: *p*=0.21), age (discovery: *p*=0.41; PPMI: *p*=0.45), or education (discovery: *p*=0.57; PPMI: *p*=0.26) in either cohort. Demographic information for participants in both cohorts is summarized in Table 1.

In the discovery cohort, results revealed a reduction in SNpc MD (Parkinson's: $4.40 \times 10^{-4}$ mm$^2$/s ± $7.3 \times 10^{-5}$ mm$^2$/s; control: $4.72 \times 10^{-4}$ mm$^2$/s ± $4.8 \times 10^{-5}$ mm$^2$/s; *P*=0.017; *t*=2.0) and SNpc RD (Parkinson's: $3.33 \times 10^{-4}$ mm$^2$/s ± $6.6 \times 10^{-5}$ mm$^2$/s; control: $3.61 \times 10^{-4}$ mm$^2$/s ± $5.1 \times 10^{-5}$ mm$^2$/s; *P*=0.022; *t*=2.0) in the Parkinson's disease group relative to the control group. An increase in SNpc FA was seen in the Parkinson's disease group as compared to the control group (Parkinson's: 0.44 ± 0.05; control: 0.42 ± 0.03; *P*=0.044; *t*=-2.0). No significant difference was observed in SNpc AD (Parkinson's: $6.60 \times 10^{-4}$ mm$^2$/s ± $9.4 \times 10^{-5}$ mm$^2$/s; control: $6.92 \times 10^{-4}$ mm$^2$/s ± $8.2 \times 10^{-5}$ mm$^2$/s; *P*=0.065; *t*=1.5) between groups. Bi-compartment analysis revealed an increase in the SNpc free water compartment of Parkinson's patients as compared to controls (Parkinson's: 0.27 ± 0.04; control: 0.23 ± 0.05; *P*=0.006; *t*=2.6). An increase in mean SNpc $R_2^*$ was seen in the Parkinson's disease group as compared to the control group (Parkinson's: 31.3 s$^{-1}$ ± 4.4 s$^{-1}$; control: 26.0 s$^{-1}$ ± 3.0 s$^{-1}$; *P*<0.001; *t*=5.429).

In the PPMI cohort, no significant differences were seen in SNpc FA (Parkinson's: 0.49 ± 0.06; control: 0.49 ± 0.05; *P*=0.371; *t*=0.329), SNpc MD (Parkinson's: $6.51 \times 10^{-4}$ mm$^2$/s ± $1.11 \times 10^{-4}$ mm$^2$/s; control: $6.40 \times 10^{-4}$ mm$^2$/s ± $1.24 \times 10^{-4}$ mm$^2$/s; *P*=0.330; *t*=-0.441), SNpc RD (Parkinson's: $4.86 \times 10^{-4}$ mm$^2$/s ± $1.20 \times 10^{-4}$ mm$^2$/s; control: $4.74 \times 10^{-4}$ mm$^2$/s ± $1.04 \times 10^{-4}$ mm$^2$/s; *P*=0.295; *t*=-0.541), or SNpc AD (Parkinson's: $9.41 \times 10^{-4}$ mm$^2$/s ± $1.43 \times 10^{-4}$ mm$^2$/s; control: $9.72 \times 10^{-4}$ mm$^2$/s ± $1.34 \times 10^{-4}$ mm$^2$/s; *P*=0.122; *t*=1.171). An increase in SNpc FW (Parkinson's: 0.19 ± 0.06; control: 0.16 ± 0.04; *P*=0.01; *t*=-2.379) and SNpc $R_2$ (Parkinson's: 11.9 s$^{-1}$ ± 1.0 s$^{-1}$; control: 11.6 s$^{-1}$ ± 0.8 s$^{-1}$; *P*=0.023; *t*=-2.022) was seen in the Parkinson's disease group compared to the control group. Group comparisons from both cohorts are summarized in Figure 3.

The effect of Parkinson's disease-related iron deposition on diffusion signal in SNpc was assessed by correlating SNpc diffusion indices (MD, RD, AD, and free water measures) with SNpc iron measures in both cohorts. SNpc $R_2^*$ was found to be negatively correlated with SNpc MD (*r*=-0.400; *P*=0.006; *N*=39), SNpc RD (*r*=-0.333; *P*=0.019; *N*=39), and SNpc AD (*r*=-0.426; *P*=0.003; *N*=39) in the Parkinson's disease group from the discovery cohort. Similarly, SNpc $R_2$ was found to be negatively correlated with SNpc MD (*r*=-0.387; *P*=0.003; *N*=52), SNpc RD (*r*=-0.411; *P*=0.002; *N*=52), and SNpc AD (*r*=-0.335; *P*=0.009; *N*=52) in the Parkinson's disease group from the PPMI cohort. However, no correlation was observed between iron measures and free water measures in the Parkinson's disease group of the discovery cohort (SNpc $R_2^*$ and SNpc FW: *r*=0.134; *P*=0.415; *N*=39) or in the PPMI cohort (SNpc $R_2$ and SNpc FW: *r*=-0.189; *P*=0.104; *N*=52). These correlations are shown in Figure 4. No correlations were observed between free water measures and iron measures in the control group of the discovery cohort (SNpc $R_2^*$ and SNpc FW: *r*=0.270; *P*=0.156; *N*=32) or in the PPMI cohort (SNpc $R_2$ and SNpc FW: *r*=-0.201; *P*=0.145; *N*=58).

Diffusion measures of MD, RD, and AD were found to be correlated with iron ($R_2^*$ or $R_2$) in both cohorts. Thus, to evaluate Parkinson's disease related microstructural changes without the contribution of iron, we performed a *post hoc* ANCOVA on MD, RD, and AD measures. In the discovery cohort, no significant effects were observed in the ANCOVA analysis for MD ($p=0.868$; $F=0.028$), RD ($p=0.618$; $F=0.251$), or AD ($P=0.717$; $F=0.132$). These results indicate that the group differences observed in nigral diffusivity in the discovery cohort were due to iron. No significant effects were seen in the ANCOVA analysis for MD ($P=0.328$; $F=0.965$), RD ($P=0.310$; $F=1.041$), or AD ($P=0.432$; $F=0.513$) in the PPMI cohort.

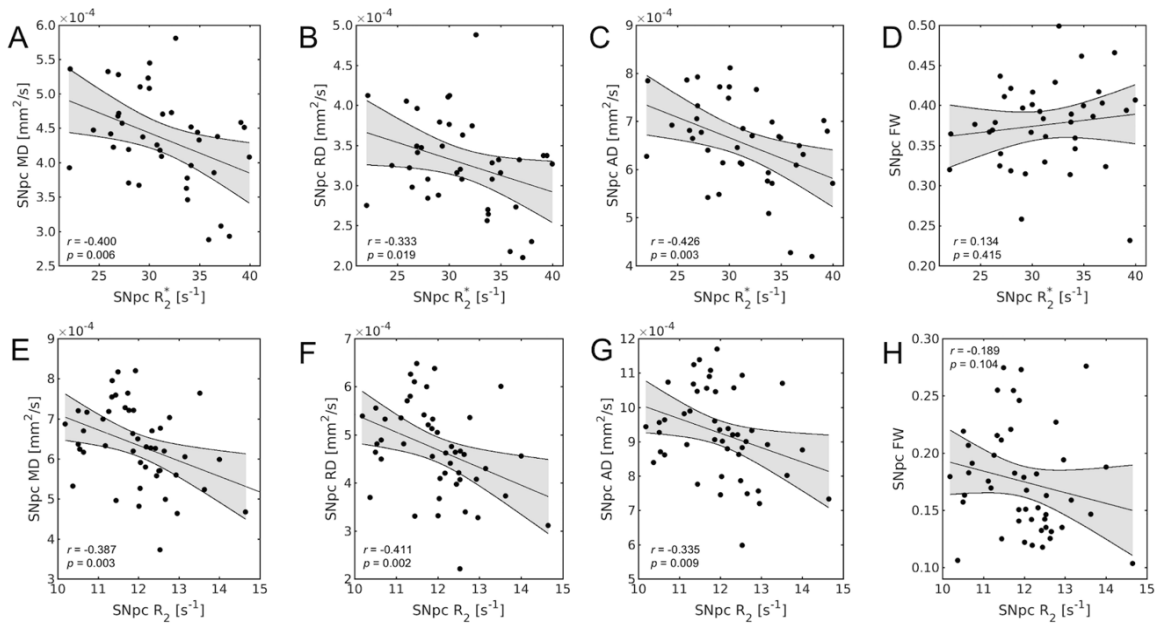

**Figure 4 The relationship between iron and diffusivity in Parkinson's disease.** Correlations between $R_2^*$ and diffusivity measures in the discovery cohort are shown in (**A-D**) and correlations between $R_2$ and diffusivity measures in the PPMI cohort are shown in (**E-H**). Significant correlations between MD, RD, and AD were seen in both cohorts while no significant correlations were observed for FW and iron in either cohort.

The Parkinson's disease group of the discovery cohort showed no correlation between ON UPDRS-II score and $R_2^*$ ($r=-0.059$; $P=0.724$; $N=39$), FW ($r=0.010$; $P=0.954$; $N=39$), FA ($r=0.116$; $P=0.483$; $N=39$), or MD ($r=-0.196$; $P=0.233$; $N=39$). Disease duration exhibited no association with $R_2^*$ ($r=0.196$; $P=0.239$; $N=39$), FW ($r=0.011$; $P=0.953$; $N=39$), FA ($r=0.239$; $P=0.143$; $N=39$), or MD ($r=-0.073$; $P=0.659$; $N=39$) in the Parkinson's disease group of the discovery cohort. In the Parkinson's disease group of the PPMI cohort, no association was seen between ON UPDRS-III score and $R_2$ ($r=-0.064$; $P=0.683$; $N=52$), FW ($r=0.121$; $P=0.428$; $N=52$), FA ($r=0.065$; $P=0.673$; $N=52$), or MD ($r=0.054$; $P=0.725$; $N=52$). No significant correlation was found between disease duration and $R_2$ ($r=-0.099$; $P=0.497$; $N=52$), FW ($r=-0.008$; $P=0.954$; $N=52$), FA ($r=-0.035$; $P=0.807$; $N=52$), or MD ($r=-0.009$; $P=0.947$; $N=52$) in the Parkinson's disease group of the PPMI cohort.

In the discovery cohort mean SNpc $R_2^*$ outperformed SNpc diffusion indices as a diagnostic imaging marker. The area under the ROC curve (AUC) for mean SNpc $R_2^*$ was 0.813 (standard error (SE)=0.056; 95% confidence interval (CI): 0.704–0.922; $P<10^{-4}$). The AUC for mean SNpc FW was 0.600 (SE=0.068; 95% CI: 0.527–0.792; $P=0.025$). In the PPMI cohort SNpc $R_2$ and SNpc FW performed similarly as diagnostic imaging markers. The AUC for SNpc $R_2$ was 0.609 (SE=0.06; 95% CI: 0.497–0.721; $P=0.05$) and SNpc FW was 0.611 (SE=0.05; 95% CI: 0.502–0.719; $P=0.05$). ROC curves for both cohorts are shown in Figure 5.

## 4. Discussion

This study examines Parkinson's disease-related changes in SNpc microstructure and iron content in two cohorts with similar clinical characteristics. Parkinson's disease patients experience a loss of melanized neurons in SNpc[2, 3] and elevated levels of non-heme ferric iron ($Fe^{3+}$) are observed alongside the loss of melanized neurons.[4-6] Elevated iron levels should impact MRI images by causing larger transverse relaxation rates while reductions in melanized neurons should cause increases in MRI measures sensitive to free water, as this compartment would be increased in areas of neuronal loss. In line with histological findings, we observed increases in iron measures and free water measures in SNpc the Parkinson's disease group of both cohorts. However, diffusivity measures from the single-compartment DTI model (FA, MD, RD) were found to be inconsistent Parkinson's disease diagnostic markers, with the discovery cohort showing altered SNpc diffusivity in the Parkinson's disease group, while no group differences were seen with these DTI markers in SNpc in the PPMI cohort. All single-compartment diffusion markers (FA, MD, RD, and AD) were found to be significantly correlated with iron, whereas the two-compartment model diffusion measure, FW, did not correlate significantly with iron. Together, these results suggest that variability in tissue iron levels may impact diffusion MRI results with single-compartment DTI measures but not FW from the bi-compartment model.

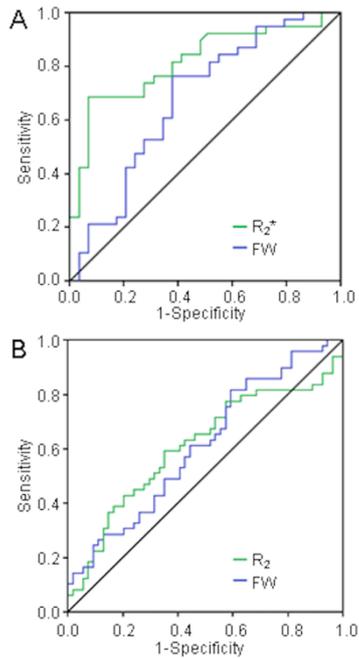

**Figure 5 Receiver operator characteristic analyses of nigral iron and free water**. (**A**) Receiver operator characteristic curves for the discovery cohort. (**B**) Receiver operator characteristic curves for the PPMI cohort.

Neuromelanin granules in substantia nigra chelate iron.[54, 55] Sequestered iron may be released when neuromelanin granules are phagocytosed and degraded after the loss of melanized neurons. Increases in transverse relaxation rates were seen in SNpc of the discovery cohort ($R_2^*$) and the PPMI cohort ($R_2$), indicating iron is being deposited in SNpc of Parkinson's disease patients. These results agree with earlier studies that found an increase in mean SNpc $R_2^*$ of Parkinson's disease patients[52, 56] and those reporting an increase in mean nigral transverse relaxation rates in regions drawn in the $T_2$-weighted substantia nigra.[11, 12, 27-33]

Measures sensitive to the free water compartment in SNpc were found to increase in the Parkinson's disease group of both cohorts. These results are consistent with earlier work reporting increased nigral free water of Parkinson's disease patients.[20, 21, 23, 25, 57, 58] The free water compartment represents extracellular water molecules (i.e. molecules unhindered by cellular environment) within a voxel.[46, 59] Increases in the free water compartment are generally interpreted as a reduction of of SNpc neuronal density.[21, 23] Since the ROI used to define SNpc was derived from neuromelanin-sensitive images, the increase in the free water compartment of the Parkinson's disease group of both cohorts may reflect a loss of melanized neurons in SNpc.

An increase of the free water compartment should yield an increase in diffusivity from the single-compartment model. However, a reduction in SNpc diffusivity was observed in the Parkinson's disease group of the discovery cohort and this agrees with an earlier study reporting reduced SNpc diffusivity in a similar SNpc ROI.[15] This reduction in diffusivity may be due to the influence of iron as higher iron measures have been observed to negatively bias nigral diffusivity in older adults[34] or striatal diffusivity in older adults[35] or patients with Huntington's disease.[53] Local magnetic field gradients from iron deposits produce cross terms with diffusion encoding gradients and reduce the apparent diffusion coefficient.[60, 61] In agreement with this model, we observed negative correlations between mean SNpc $R_2^*$ and SNpc diffusivity in the Parkinson's disease group of the discovery cohort.

Mean SNpc single-compartment diffusivity measures in the Parkinson's disease group of the PPMI cohort were negatively correlated with mean SNpc $R_2$, but no group differences in single-compartment diffusivity measures were observed. This is in agreement with an earlier study which found no difference in nigral diffusivity measures between Parkinson's disease and control groups.[62] The lack of a disease effect for single-compartment diffusion markers in these studies may be attributed to two competing influences: a reduction in melanized neurons, which tends to drive diffusivity up and FA down, and iron deposition, which tends to drive diffusivity down and FA up. In some cases, these competing effects may offset and reduce the effect size. These competing effects may partially explain variability observed in earlier diffusion studies examining nigral DTI metrics where studies have reported lower nigral FA[10-13, 15] or no difference in nigral FA.[8, 16-19]

The ROC analysis found AUC of nigral $R_2^*$ to be comparable to previously published diagnostic markers examining nigral tissue composition [63-68], which have AUCs between 0.7 and 0.9. However, AUC for nigral FW in both cohorts was significantly below AUCs reported in other studies.[23, 69] The discrepancy in performance of FW imaging markers may be related to ROI selection. We used the entire SNpc was used as a ROI in the AUC analysis while earlier studies reported AUC in SNpc subregions with ROIs placed in the posterior SNpc.[20, 21, 23, 70] Nigrosome-1, the SNpc subregion that experiences the greatest loss of melanized neurons, is located in the posterior portion of SNpc[49] and posterior SNpc ROIs likely capture degeneration in nigrosome-1 whereas the entire SNpc ROI will contain nigrosome-1 as well as other regions that lose fewer melanized neurons.

The study has several caveats. First, only a subset of the PPMI cohort was used in the diffusion and iron analyses. This was due to inconsistent $R_2$ and DTI scan parameters at several imaging sites. Second, Parkinson's disease-related iron deposition is expected to reduce signal-to-noise ratios in SNpc. This deposition should increase noise and may corrupt diffusion measures in the discovery and PPMI cohorts.

Reduced signal to noise ratios will positively bias AD and FA while negatively biasing RD.[71] We speculate that the effect of noise is minimal in both cohorts since correlations of similar strength are seen between iron metrics and both RD and AD. Third, monopolar diffusion encoding gradients were applied in both cohorts. Bipolar diffusion encoding gradients may be less sensitive to iron deposits.[72] However, a recent study found $R_2*$ negatively biases diffusivity[34] and additional work is needed to fully assess the contribution of iron on diffusivity in grey matter structures.

In this work, Parkinson's disease-related changes in SNpc microstructure and iron content were examined in two cohorts with similar clinical characteristics. Measures sensitive to FW and iron content were found to increase in SNpc of the Parkinson's disease group in both cohorts. However, diffusion markers derived from the single-compartment model (i.e. MD, RD, AD, and FA) were not replicated across cohorts. The variability of metrics derived from the single-compartment model may be attributed to competing influences of iron content[34, 35, 53] and free water on the diffusion signal, the placement of ROIs outside SNpc,[15, 17] or a combination of these factors. In contrast to SNpc diffusivity, no association was found between SNpc FW and SNpc iron measures in either cohort. This insensitivity to iron, coupled with consistent observations of increased nigral free water from this study and earlier studies,[20, 21, 23, 25, 57, 58] suggests the free water compartment of bi-compartment diffusion models should be used in lieu of diffusivity measures derived from the single-compartment model to study SNpc.

## Acknowledgements

This work was supported 1K23NS105944-01A1 (Huddleston) from the National Institutes of Health/National Institute of Neurological Diseases and Stroke, the Michael J. Fox Foundation grants MJFF-010556 and MJFF-010854 (Huddleston, Langley and Hu), and the American Parkinson's Disease Foundation Center for Advanced Research at Emory University (Huddleston). PPMI – a public-private partnership – is funded by the Michael J. Fox Foundation for Parkinson's Research and funding partners, including [list the full names of all of the PPMI funding partners found at www.ppmi-info.org/fundingpartners].

## Competing Interests

The authors report no competing interests.